\documentclass[onecolumn]{aastex6}

\shorttitle{Period Analysis from Multiband Data}
\shortauthors{A. Saha \& A.K.~Vivas}

\begin{document}


\title{A Hybrid Algorithm for Period Analysis from Multi-band Data with Sparse and Irregular Sampling for Arbitrary Light Curve Shapes}


\author{Abhijit Saha}
\affil{National Optical Astronomy Observatories, 950 N. Cherry Ave, Tucson, AZ 85719}
\email{saha@noao.edu}

\author{A. Katherina Vivas}
\affil{Cerro Tololo Inter-American Observatory, National Optical Astronomy Observatories, Casilla 603, La Serena, Chile}
\email{kvivas@ctio.noao.edu}

\begin{abstract}
Ongoing and future surveys with repeat imaging in multiple bands are producing (or will produce) time-spaced measurements of brightness, resulting in the identification of large numbers of variable sources in the sky.  A large fraction of these are periodic variables: compilations of these are of scientific interest for a variety of purposes.   Unavoidably, the data-sets from many such surveys not only have sparse sampling, but also have embedded frequencies in the observing cadence that beat against 
the natural periodicities of any object under investigation. Such limitations can make period determination ambiguous and uncertain.
For multi-band data sets with asynchronous measurements in multiple pass-bands, we want to maximally utilize the information on periodicity in a manner that is agnostic of differences in the light curve shapes across the different channels.  Given large volumes of data, computational efficiency is also at a premium. This paper develops and presents a computationally 
economic method for determining periodicity which combines the results from two different classes of period determination algorithms. The underlying principles are illustrated through examples.  The effectiveness of this approach for combining asynchronously sampled measurements in multiple  observables that share an underlying fundamental frequency is also demonstrated. 
 
\end{abstract}


\section{Introduction}
\label{sec:intro}
A common problem in time-domain astronomy is the identification and
determination of periodicity in variable phenomena. The available data
are all too often sparsely sampled, with irregular intervals taken
with cadences which themselves have embedded periodicities that beat
against any periodicity and harmonics present in the variable phenomena
being investigated.  In the face of this, astronomers have
historically been quite resourceful at coaxing out periodicites from
data that are beset with aliasing and noise.
A comprehensive review of the various techniques (which have grown in
complexity as computational power has increased) falls outside the
scope of this paper. The reader can find an admirable recent review in
\citet{graham13}.  The deluge of variability data from future surveys
such as the Large Synoptic Survey Telescope (LSST) will have irregular
and sparse sampling, measurements in multiple passbands that are
separated in time from band to band, and embedded periodicities in the
cadence (diurnal, lunar and annual cycles as well as from chosen
cadences). This paper presents a procedure that seeks to combine the
most computationally fast existing procedures to produce a robust
technique for period determination in the face of such sub-optimal
data.

The last decade has produced several time-domain photometric surveys,
e.g. the Catalina Real-Time Transient Survey \citep{Drake09} , the
Palomar Transient Factory \citep{Law09, Rau09}, Pan-STARRS
\citep{Chambers16}, and Kepler \citep{Koch10} to name a few, which
have added to the legacy of MACHO \citep{Alcock92, Alcock96} , OGLE
\citep{Udalski93} and ASAS \citep{Pojmanski02}. All of these are
slated to be dwarfed by the 10 year projected survey with Large
Synoptic Survey Telescope (LSST) which is currently under
construction. While many of the above named projects were exclusively
time-domain surveys, the broader mission of the LSST survey will
result in collecting brightness measures for a given source in the sky
at $\sim 1000$ epochs spread over 10 years.  These $\sim 1000$ visits
will be spread over 6 different passbands, so that there will be
between $ 100 {~\rm to~} 200 $ total samples in any given passband.
Thus at the end of the first year's observing season, we can expect
there to be between 10 to 20 epochs for a given object on average in
any given passband. The observations in different bands will not be
near simultaneous, so in principle there will be up to 100 epochs in
total spread over all passbands.  Even though we can expect that the
observing cadence might ensure that each part of the sky gets visited
oftener than the average cadence at {\emph some} times over the 10
year survey, early determination of light curves will surely be at a
premium for any and all variable sources. An efficacious method for
gleaning periodicities early in the survey from such sparse sampling
over varied passbands is thus desirable.

The authors of this paper have recently been engaged in a time-domain
survey of the Galactic bulge area with the DECam imager
\citep{flaugher15} on the 4-m Blanco telescope at CTIO. Observations
have been made of 6 select fields, in 5 different passbands.  Although
the cadence was optimized for characterizing RR~Lyrae stars (the
primary purpose of the survey), the sampling remains sparse, and with
embedded periodicities in the sampling pattern (diurnal, lunar, and
annual cycles).  Also, given the multiple passbands, the data-set
provides a good test bed for examining how partial information from
the individual bands can be combined to infer periodicity information
that is common to all. The methodology presented in this paper was
developed to improve the efficiency of period determination and
identification of desired variable stars from over 20,000 objects
flagged as variable.

Existing methods for period determination can be broadly binned into
two classes. In the first, the data are folded by a set of trial
periods, so that each observation at time $t$ with measured magnitude
$m$ can be assigned a phase $\phi$.  We then look for order in this
$(\phi, m)$ plane, such that at the correct $P$, the observations
fall along a narrow locus that spans the range of $\phi$.  No prior
assumption need be made of the light curve shape, and the approach is
shape agnostic. Differences between methods of this ilk, which we
label as the `Phase Folding Method' (PFM), lie in how they go about
measuring the scatter of the observations in the $(\phi, m)$ plane.
The second class of methods uses harmonic analysis (HA): Fourier and
wavelet based methods fall into this category. A specific
implementation is to invoke the properties of the Fourier series where
the orthonormality of Fourier coefficients for a harmonic progression
of putative periods can be used to uniquely describe any periodic
signal of any shape. There are two notable issues: 1) deriving the
Fourier spectrum requires regular sampling, which hardly any
astronomical data-sets provide, so a proxy for the Fourier spectrum is
often used; and 2) the shape of the light curve is manifest in the
Fourier power-spectrum as the relative strength and phases at
different frequencies.  We do not consider template fitting methods
here: while they are very useful if you know the shape of the light
curve, our interest is in the more general case where the shape is
unconstrained.  In this paper a representative method from each of the
PFM and harmonic analysis classes has been chosen.  The selection
favored algorithms that are computationally straightforward and
economical, and which, in the authors experience, have been gainfully
employed across a variety of period searching projects over the past
several decades, thus authenticating their robustness.  While within
each class one particular implementation may better another
marginally, this paper shows that substantive gains are to be had by
combining results across the two classes since they provide
complementary information on the true period.

In section \S~\ref{sec:methods}, we show how these methods respond to
different kinds of variable signals, with differing sampling density,
cadence, and noise. A new period discriminator (that combines
information from the PFM and harmonic analysis classes) is introduced,
and its efficacy is demonstrated. The two chosen methods from each of
the classes mentioned above are described and applied to archetypal
examples spanning different shapes of light curves, different sampling
patterns, sparseness, and noise, gradually increasing in complexity.
The pedagogical walk through the examples illuminates the strengths
and weaknesses of the two classes of methods. A statistic that
combines the analysis from both classes is introduced, and its
efficacy is illustrated through the examples considered.  In
\S~\ref{sec:practical}, key constraints that govern how densely spaced
the search for periods or frequencies must be are discussed.  In
\S~\ref{sec:real_data} a real multi-band data-set for an RR~Lyrae star
is examined, and it is shown how the new discriminant for the
individual bands can be combined to reinforce information across
bands.  \S~\ref{sec:confthresh} discusses how confidence estimates are
generated for the new discriminant.  \S~\ref{sec:LSSTsim} applies the
methodology for an observational cadence simulated for LSST, to show
that it works for asynchronous measurements made in the different
pass-bands.  Computer code in the IDL language is provided via {\it
  github}, as described in the Appendix.  The problem of combining
asynchronous measurements from different channels (e.g. multiple
passbands) to derive the underlying common fundamental period has also
been addressed by \citet{vanderP15} and by \citet{mondrik15}, both
wholly within the HA approach.
We comment on how the method presented in this paper differs from
theirs in the concluding discussion in \S~\ref{sec:Conc}

\section{Period Analysis from a single observable}
\label{sec:methods}

It is useful first to assign specific meaning to terms that are used
in this paper.  The term `time-sequence' will refer to the data-set
being analyzed, i.e. a set of data-points with observation time $t$,
and the time dependent variable, such as flux, magnitude, velocity,
etc, is designated by $m$.  We are interested here in periodic
behavior: when a time-sequence is folded by a period $P$ (whether
assumed or real), it produces a `light curve' where the data-points
$m$ are assigned a phase $\phi$.

\begin{equation}
\label{eqn:phase}
 \phi =  t/P  ~~modulo~~ 1
\end{equation}

We use the term `periodogram' for the period/frequency spectrum of \emph {any} 
periodicity indicating metric.
For a given astrophysical object, we may have more than one observed
quantity that is varying in time, e.g. magnitude and velocity, or
magnitudes in different passbands.  These share a common periodicity,
although they may have quite different wave-forms in these different
apparitions. Said another way, all of these quantities share a \emph
  {fundamental period}, but differ in the relative strengths of the
higher harmonics.  We address here the problem of how best to find the
fundamental period, and in the methodology of combining information
from multiple observed quantities that share it.

\subsection{The Phase Folding Method: Phase Dispersion Minimization and the Lafler-Kinman approach}
\label{sec:lk}

The phase dispersion minimization \citep[PDM,][]{stellingwerf78}
method is based on the idea that if the time-sequence data are folded
by the correct period $P$, the resulting light curve shows minimal
scatter within all bins in phase.  The sum of the scatter (or net
dispersion) from all bins is obtained for each trial period, and the
period for which this sum is the smallest is notionally the correct
period.  The binning need not be done to keep equal bin sizes: for
sparse data, where phase coverage (which changes with assumed period)
is likely to be uneven, it is possible to use an adaptive binning
strategy using all possible bins, each with $n$ points. A widely used
implementation of this technique which precedes the
\citet{stellingwerf78} generalized PDM method, is the Lafler-Kinman
\citep[LK,][]{lafler67} procedure, which is in effect a special case
of PDM, with adaptive binning and $n=2$.  The choice of minimal $n$ is
particularly suitable for sparse data, since it keeps the bins as
narrow as the data will allow. It is chosen here as the representative
of the PFM class because it is computationally the most economical,
conceptually the simplest, and has served the senior author of this paper well
for over three decades.

The Lafler-Kinman statistic $\Theta$ at a given period $P$ is given by:
\begin{equation}
\Theta  =  \frac{  \sum_{i=1}^{N} (m_{i} - m_{i-1})^{2}  } {  \sum_{i=1}^{N} (m_{i} - \overline{m})^{2} }
 \end{equation}

 where the $N$ measurements $m_{j}$ at times $t_{j}$ have been folded
 with period $P$ to produce phases $\phi_{j}$ in accordance with
 eqn.~\ref{eqn:phase} and sorted in ascending order of $\phi$, to get
 ordered pairs $(\phi_{i}, m_{i})$. In addition, in the phase sorted
 arrays, $(\phi_{0}, m_{0})$ is set equal to $(\phi_{N}, m_{N})$, which
 completes the phase cycle. $\overline{m}$ is the mean of all $m_{i}$.
 This formula assumes equal weighting of all points.  Let
\begin{equation}
 w_{i} = \frac {1} { (\sigma_{i}^2 + \sigma_{i-1}^2)}
\end{equation}

where $\sigma_{i}$ is the uncertainty in $m_{i}$.  The Lafler-Kinman can then be modified as:
\begin{equation}
\Theta  =  \frac{  \sum_{i=1}^{N}  w_{i}(m_{i} - m_{i-1})^{2}  } {  \sum_{i=1}^{N} (m_{i} - \overline{m})^{2}  .  \sum_{i=1}^{N} w_{i} }
\end{equation}
where $\overline{m}$ is correspondingly the weighted mean over $m_{i}$.

As for any PFM class method, a strength of the LK approach is that it
makes no assumption about the shape of the light curve.  A weakness is
that as data get sparse, the gaps in successive phase in the folded
trial light curves can become very different from one period to
another, resulting in periodograms (period vs. net dispersion) with
increasing false structure which can interfere with the
identification of the true period.  Thus in addition to aliasing,
there are spurious minima produced in the periodogram that are introduced 
by the paucity of data.  For
practical applications, it is important to consider limitations on the
step-size between successive trial periods, so that the true period is
not missed: details are discussed later in \S~\ref{sec:practical}.

\subsection{Harmonic Analysis and the Lomb-Scargle approach}

Any periodic function with arbitrary shape that is piece-wise
continuous with a finite number of finite discontinuities (Dirichlet's
conditions) can be represented by a Fourier series with frequencies
that are integer multiples of the fundamental frequency.  Thus period
analysis of evenly spaced time-series data is straightforward.  For
uneven spacing, to perform a strict Fourier analysis, the data must be
re-sampled evenly.  The re-sampling process can be computationally
expensive, and can introduce significant specious structure in the
Fourier power spectrum.  Alternatively, the amplitude of a fitted
sine(/cosine)-wave at a given frequency can be used to glean the
Fourier power at that frequency, and a periodogram can thus be constructed 
over desired frequencies, thereby eliminating the need to re-sample the
data.  Credit for implementing the latter approach to astronomy
applications belong to \citet{lomb76} and \citet{scargle82}. The
eponymous Lomb-Scargle algorithm is available in various coding
languages.

A Fourier transform (FT) or a Lomb-Scargle (LS)
periodogram\footnote{the LS periodogram is considered henceforth as
  having the same properties of an FT, which it does if the
  frequencies sampled are part of a harmonic series} contains
information not only of the fundamental period, but also of the
waveform shape, which appear as sub-dominant peaks in the periodogram.
With well and plentifully sampled data, this can be an advantage, but
this additional signal can ironically also be a source of confusion
when the data are sparse and sampled with embedded periodicities that
produce aliasing.  If the signal being sampled is a pure sine wave,
ideal sampling in the absence of noise 
will produce a single peak in the periodogram. However,
missing data has the effect of redistributing power into alias
frequencies (and phases): essentially a convolution of the true
frequency peak with a frequency dependent function that represents the
actual sampling (often called the `window function').  To see this,
consider a signal $U(t^{\prime})$ (true time-series), which is measured 
at discrete times $\tau_{k}$.  The available data can be
re-sampled into an even time series $S(t_{i})$, where

\begin{equation}\label{eqn:samp_window}
S(t_{i}) = U(t_{i}) . W(t_{i})
\end{equation}

where $W(t_{i}) = 1$ when the time bin $t_{i}$ contains any of
$\tau_{k}$, and $0$ otherwise.  The Fourier domain manifestation of
the sampled signal $S$ is then the convolution of the Fourier domain
representations of the `true' signal $U$ and the `window function' $W$.

For non-sinusoidal variation, the Fourier coefficients representing
higher harmonics are also convolved by the window-function,
potentially creating confusion.  Noise is an additional complication,
since a full harmonic analysis will also seek to fit individual
excursions due to noise, generating a Fourier spectrum that has power
at frequencies other than the fundamental frequency or its harmonics.
Thus a periodogram in general contains components other than those
from the source due to noise and missing data.

In the following subsections we examine a set of simple examples of
how the LK and the LS methods react to different waveforms, both in
the well sampled and sparsely sampled instances, as well as when there
is periodicity in the sampling windows.  The examples illustrate the
ideas mentioned above. We also see how the differences in how these
methods respond can be used to our advantage: that the simultaneous
application of both methods can be used to mitigate confusion that
each of them generates singly.  An ensuing hybrid statistic that
reinforces information about the fundamental period/frequency from
independent sampling of multiple quantities is developed and
demonstrated for real world data.

\subsection{Comparison of Lomb-Scargle and Lafler-Kinman methods in different situations}
\label{sec:FvsLK}

\begin{figure}[htb!]
\centering
\epsscale{0.85}
\plotone{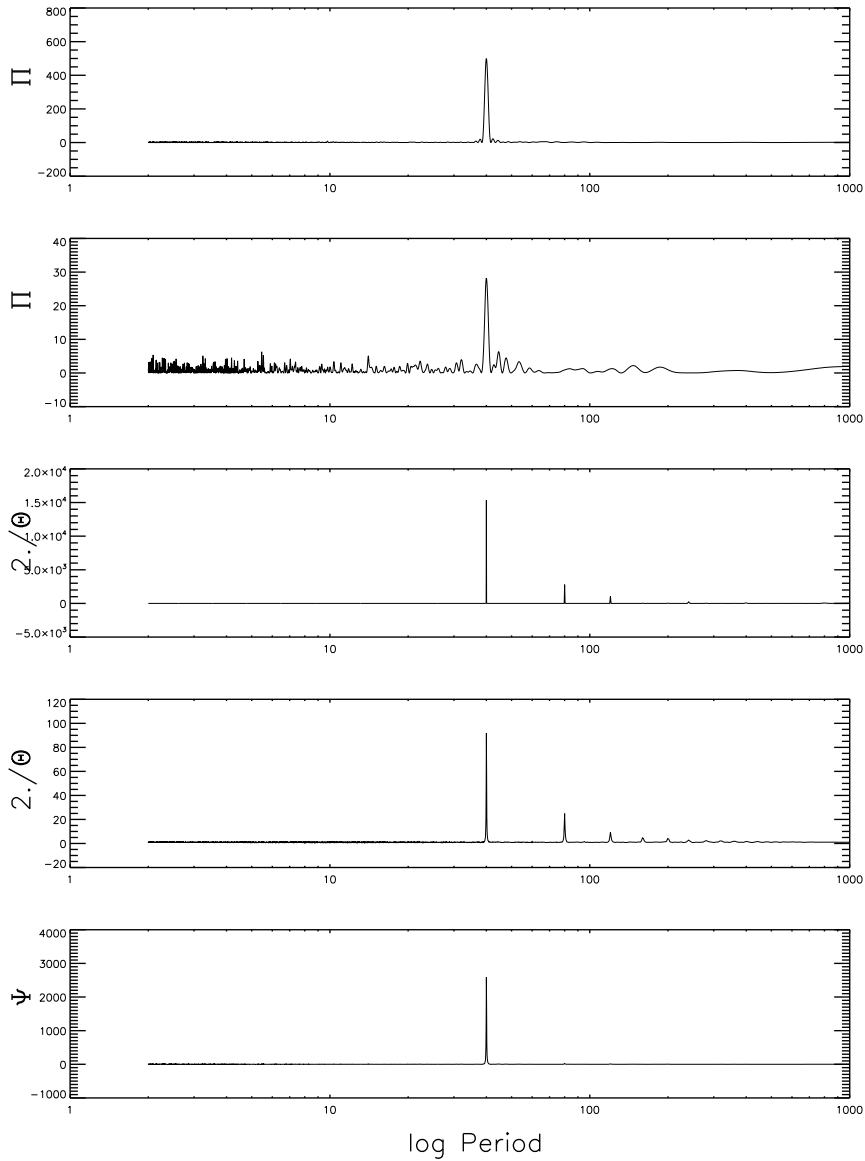}
\caption{Top panel shows the Lomb-Scargle (LS) power spectrum (plotted
  vs. log period) for a repeating sine wave with period of 40 units.
  The data are sampled 1000 times randomly, but distributed uniformly,
  over a duration of 1000 units.  A single strong peak at 40 units is
  seen as expected. The second panel shows what happens when only 5\%
  of the original samples (picked at random) are used.  The third
  panel shows the reciprocal of the Lafler-Kinman (LK) statistic
  $\Theta$ for the same data: which produces sub-harmonic features.
  The fourth panel shows the $\Theta$ statistic for the same 5\%
  sub-sample of the data. The bottom panel shows the combined
  statistic $\Psi$ for the {\it sparsely} sampled data. See
  \S~\ref{sec:FvsLK} for details and explanation.}
\label{fig:sine_random} 
\end{figure}

A few elementary examples best help to contrast and compare the
relative strengths and weaknesses of the two methods. First consider a
sine wave signal with a period $P_{0}$ of 40 units, sampled over a
total duration of 1000 units.  The data are sampled at random
instances with mean spacing of one unit, totaling 1000 samples.
The top panel of Fig.~\ref{fig:sine_random} shows the resulting LS
power ($\Pi$) spectrum, with a strong peak at the 40 unit period of
the signal.  The LS analysis was done using a harmonic series in
period (arithmetic series in frequency) so that the Fourier
coefficients at each sampled frequency are mutually orthogonal (more
on the details of choosing intervals are in \S~\ref{sec:practical}).
The next panel shows what happens when only 5\% of the original data,
picked at random, are used. The resulting periodogram has power from
the 40 unit peak re-distributed (random periodicities present in the
sampling window due to the drastic reduction in sampling).  The third
panel shows the inverse (actually $2 / \Theta$, since for random
un-periodic variation $\Theta$ tends to 2) of the Lafler -Kinman
statistic $\Theta$ of the full data set described here - so instead of
the traditional minima in $\Theta$, we look for peaks in $2 / \Theta$
to be commensurate with $\Pi$ from the LS algorithm.  It does
not do quite as well, because in addition to the expected peak in $2./
\Theta$ at 40 units, there are additional peaks at sub-harmonic
frequencies, i.e. at integer multiples of the true period.  This is
because at these frequencies the `light curve' simply repeats multiple
times within the time-span of the putative period, and the LK
statistic $\Theta$ returns favorable values for those putative periods. The
fourth panel is again the LK periodogram, but when only 5\% of the
data are used.  For a pure sine-wave, clearly the FFT approach is the
one that better isolates the true period.  The final panel shows a
hybrid quantity $\Psi$ for the sub-sampled data, where $\Psi$ is
defined as:

\begin{equation}\label{eqn:psi}
\Psi = 2  \Pi / \Theta
\end{equation}

Notice how the $\Psi$ spectrum shows the true period with better
definition than either $\Pi$ or $\Theta$ in panels two and four, and
the sub-harmonics seen in $\Theta$ are completely suppressed.  We will
see in the next examples that this is an even more useful metric in
more complicated situations.

\begin{figure}[htb!]
\centering
\epsscale{0.85}
\plotone{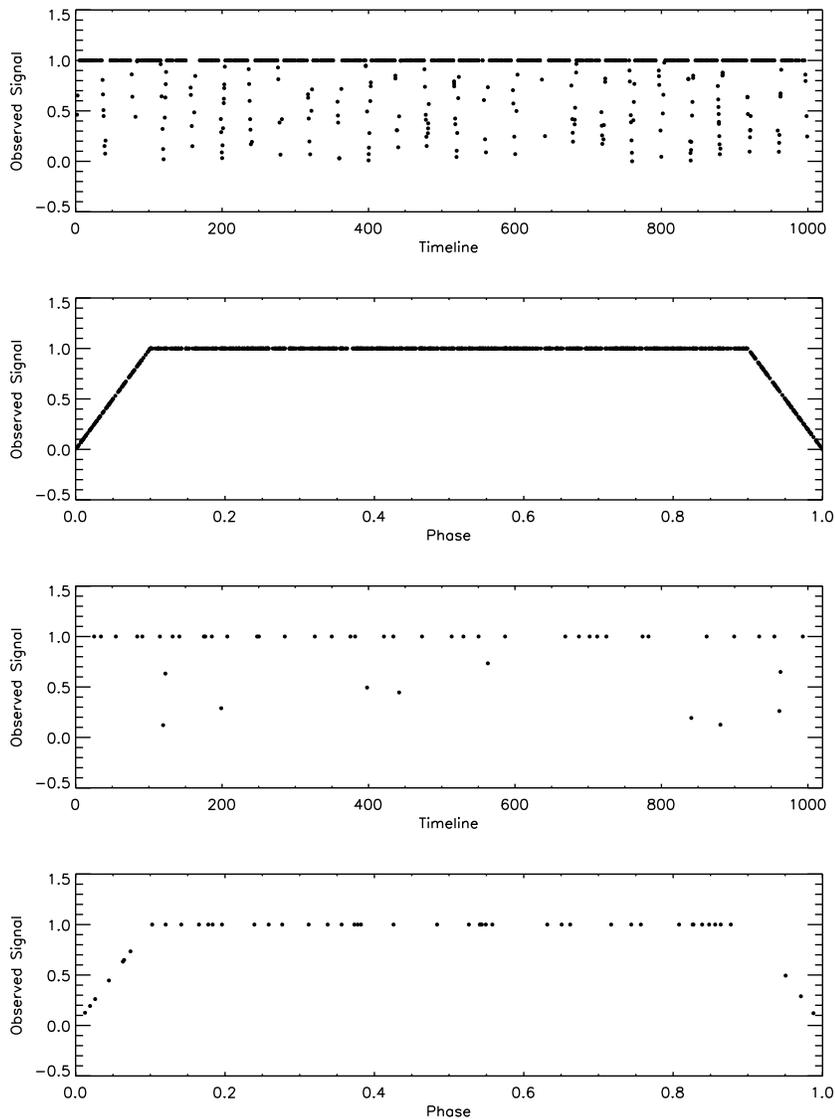}
\caption{This figure visualizes the the observation times, phases and observed values from 
 model eclipsing source described in  \S~\ref{sec:FvsLK}. The top panel shows the observation epochs and observed signal, which 
when folded by the correct period, yields the phased light curve in the second panel, which corresponds to the model described. In the third panel we see 
what happens when only 5\% of the observations (chosen at random) are available, with the the fourth panel showing the resulting phased light curve, now relatively sparsely sampled.}
\label{fig:ecl1} 
\end{figure}

Next consider the case of an eclipsing light curve, which is flat over
a 0.8 range in phase, with an eclipse modeled in the 0.2 interval of
phase as a symmetrical triangular fall and rise.  The period is again
40 units. It is sampled at random a 1000 times over a 1000 unit baseline.
The observations times and signal strength are shown in the top panel 
of Fig.~\ref{fig:ecl1},  which when folded by the correct period (40 units), 
yields the light curve in the second panel. The third panel shows the situation 
when only 5\% of the sample points are available, and the last panel shows the 
corresponding measurements when phased with the correct period. 

\begin{figure}[htb!]
\centering
\epsscale{0.85}
\plotone{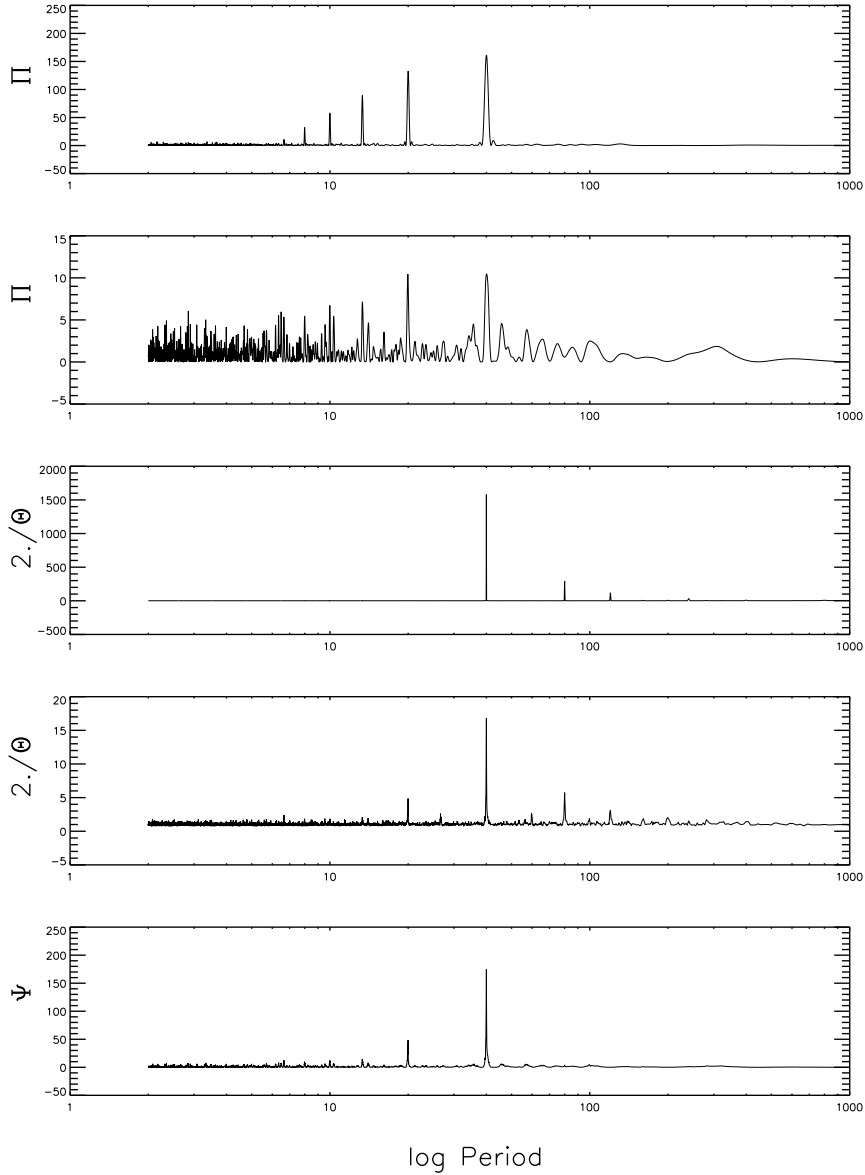}
\caption{Here shown are the results for a model light curve with an
  eclipse described in \S~\ref{sec:FvsLK} and illustrated in Fig~\ref{fig:ecl1}.  The top panel is the LS
  power spectrum for a 1000 samples taken at random times, but with
  uniform distribution across the 1000 unit time baseline.  The second
  panel is the same but with only 5\% of the sample points drawn at
  random.  The third and fourth panel are for time samples identical
  to that for panels one and two respectively, but showing the inverse
  of the LK statistic as in Fig.~\ref{fig:sine_random}.  The bottom
  panel shows the combined statistic $\Psi$ for the sparsely sampled
  data. Note the power at harmonic frequencies in the Fourier spectrum
  that are absent in the LK spectrum. See \S~\ref{sec:FvsLK} for
  details and explanation.}
\label{fig:ecl_random} 
\end{figure}

 Fig.~\ref{fig:ecl_random} shows the periodograms for these cases. Notice
that in the top panel, the higher harmonic frequencies that decompose the shape are
clearly visible as peaks at shorter periods. In the second panel,
where only 5\% of the sample points are available, there is added
confusion, as the secondary peaks become poorly resolved. The third
panel, for the LK analysis of the exact same data, shows the harmonics
very weakly if at all, since the LK method is shape agnostic.  As in
the previous case, the LK periodogram continues to show power at
sub-harmonic frequencies. The second and fourth panels show the same
analyses, but for only 5\% of the sample points taken at random: the
periodograms are similar, but with added noise-like confusion. Note
that in the final panel, $\Psi$, which combines results in the second
and fourth panels, recovers the fundamental period with the least
confusion. The LK and LS periodograms appear to each subdue the
specious features of the other.

\begin{figure}[htb!]
\centering
\epsscale{0.85}
\plotone{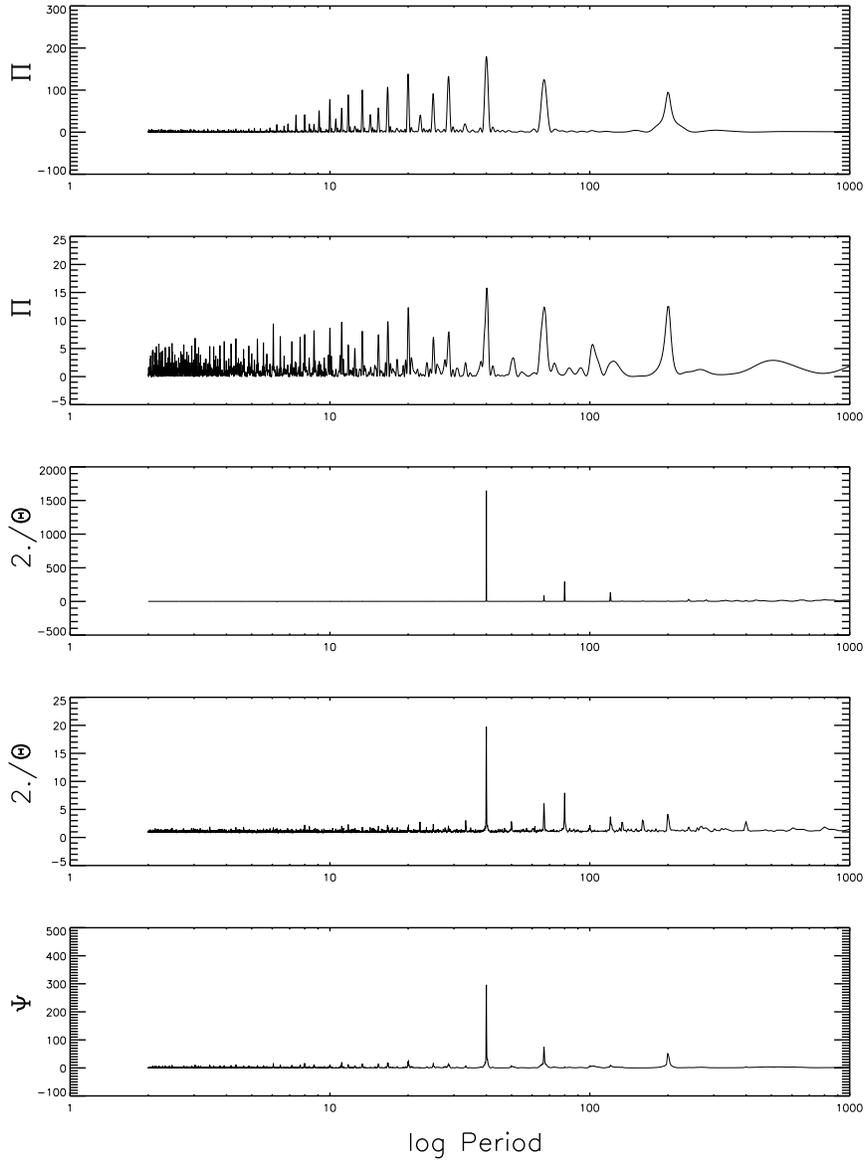}
\caption{This case differs from that in Fig.~\ref{fig:ecl_random} in
  only one way, in that the samples are taken according to a repeating
  pattern: sampling is allowed for a duration of 30 units, followed by
  a disallowed duration of 70 units. This 100 unit sampling-window
  cycle repeats periodically. Samples are taken at random times during the
  allowed intervals. In panels one and three, a total of 1000 samples
  are used, while in panels two and four, only 5\% of the sampled data
  are used. Notice how the beating of the sampling window adds peaks
  to the LS power spectrum {\it as well as} to the LK $\Theta$
  spectrum.}
\label{fig:ecl_beatrand} 
\end{figure}

Now we look at the effect of periodicities present in the sampling of
the data: consider the same eclipsing light curve with the 40 unit
period, which is again sampled at random, but with repetitive sampling
windows with period 100 units.  During each such 100 unit period,
samples can be taken only during the first 30 units (analogous to the
duration of a night).  This periodic sampling pattern beats against
the periodicity in the signal, and the results are shown in
Fig.~\ref{fig:ecl_beatrand}.  Note how additional prominent peaks have
appeared at the beat frequencies $P_{b}$, where
\begin{equation}\label{eqn:beat}
1/P_{b} = 1/P  ~\pm~ 1/P_{w}
\end{equation} 
where $P_{w}$ is the periodicity of the sampling window (in this case
100 units), and $P$ represents not only the fundamental period of the
source, but also its harmonics. This results in a `forest' of specious
peaks. Real sampling in astronomy is often infused not only with
diurnal observing windows, but also with lunation and annual cycles,
often resulting in a forest of such beat peaks. The LK spectrum also
sees the satellite alias periods with respect to the fundamental
period, but being less sensitive to intrinsic harmonics (or shape of
the light curve) is less affected at periods shorter than the true
fundamental period. We also see specious peaks on the sub-harmonic
side, since the convolving function due to embedded frequencies in
the sampling window has lobes on both sides. The $\Psi$ spectrum
reflects the best of both the LS and LK algorithms, and is the least
confused.

\begin{figure}[htb!]
\centering
\epsscale{0.85}
\plotone{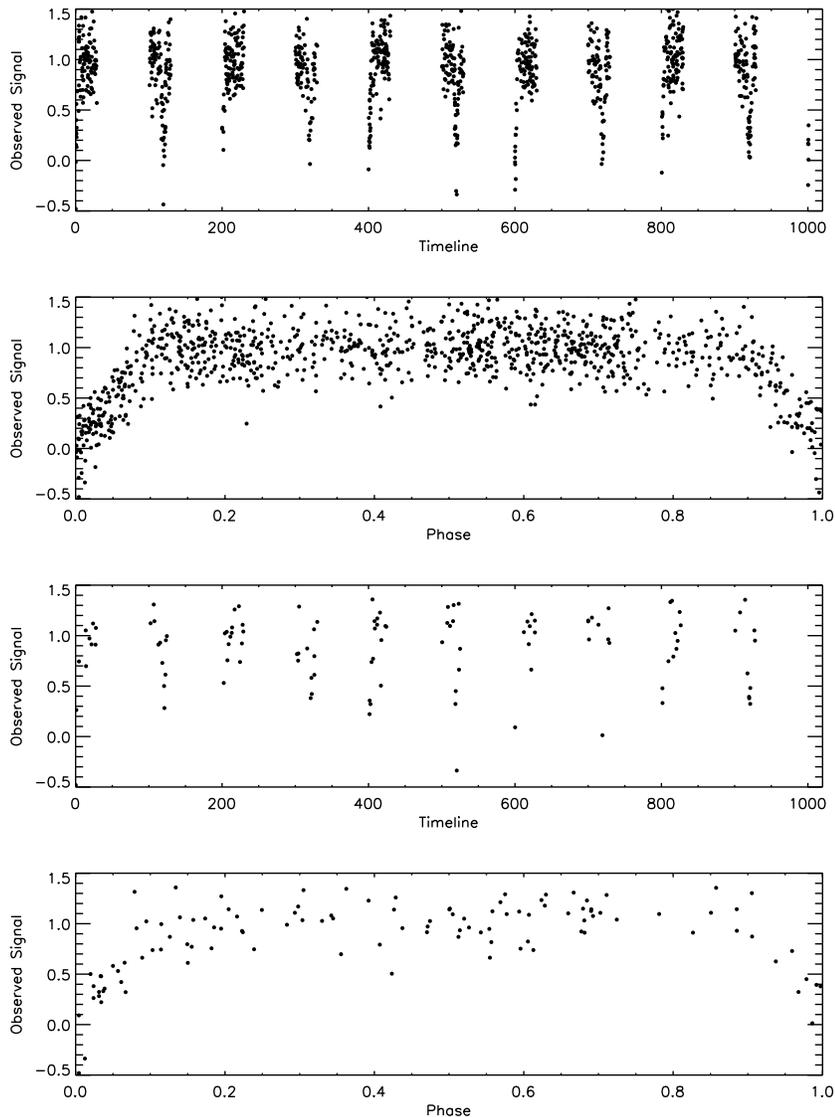}
\caption{Same as for Fig.~\ref{fig:ecl1}, but with periodic exclusion windows and with noise added, as described in \S~\ref{sec:FvsLK}. Note how in addition to the scatter from noise, the periodicity in the sampling window produces uneven sampling in phase due to beating of the object's period with the sampling period.}
\label{fig:ecl2} 
\end{figure}

In the final example, we add noise to the above case.  To each sampled
value, we add a random excursion drawn from a normal distribution with
$\sigma = 0.20$, where the amplitude of the varying signal is unity.
The net observational situation 
is illustrated in Fig.~\ref{fig:ecl2}: the top panel shows the actual observation times and 
values, and the second panel shows the light curve when the observations 
are phased with the correct period.  The third and fourth panels show the situation when 
a subset only 5\% of the measurements are available.

\begin{figure}[htb!]
\centering
\epsscale{0.85}
\plotone{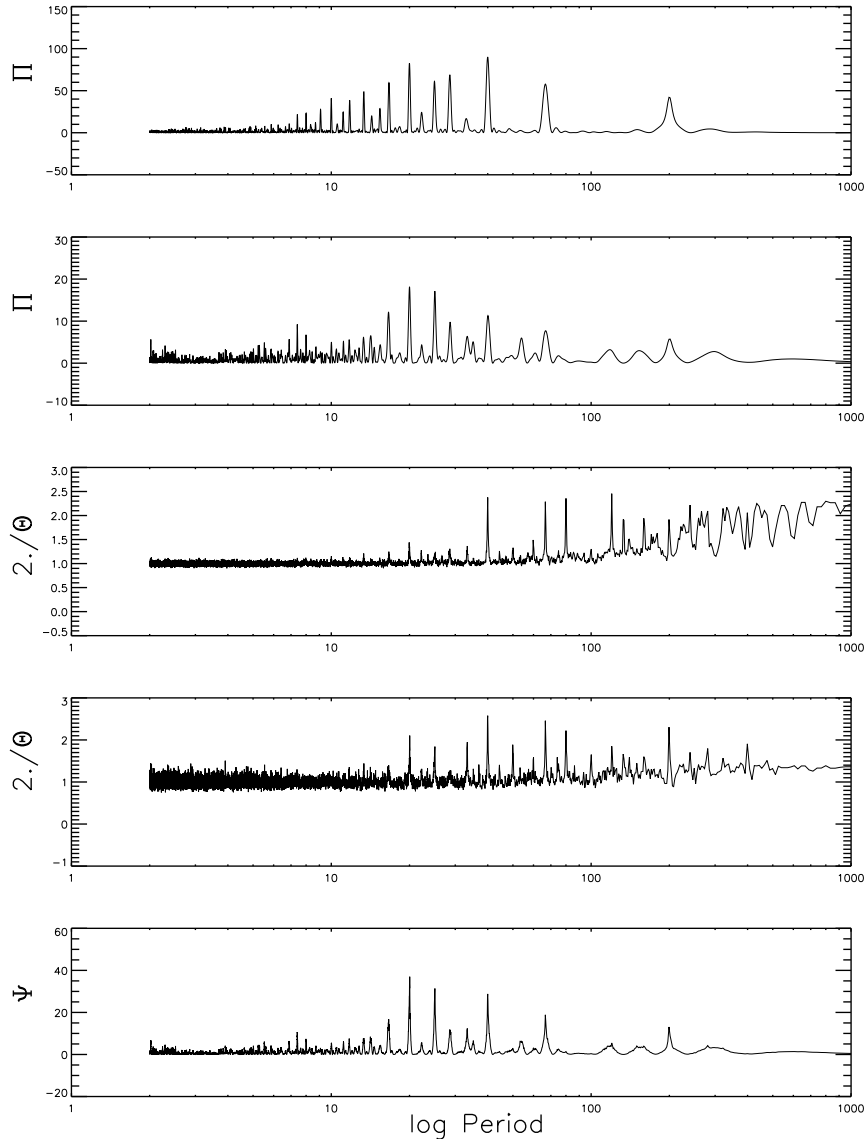}
\caption{Same as for Fig.~\ref{fig:ecl_beatrand}, but with noise added, as described in \S~\ref{sec:FvsLK}  }
\label{fig:ecl_brnoise} 
\end{figure}

The resulting periodograms with the same timing characteristics as the
previous case, are shown in Fig.~\ref{fig:ecl_brnoise}.  At this
considerably large level of noise and sparse periodic sampling (panels
two and four), the discriminating peaks are no longer unique.  In
particular it also shows that significant levels of noise are poison
to the LK algorithm, especially at sub-harmonic frequencies. But the
$\Psi$ spectrum in the final panel, which combines the information in
panels two and four, still carries the correct period among the tallest
peaks.  This supports the notion that $\Psi$ is a better period
discriminant than either $\Pi$ or $\Theta$ by themselves, in the face
of sparse irregular sampling with imposed periods of noisy data.

In \S \ref{sec:real_data}, we will study the response to a real light
curve with sparse sampling and multiple periodicities and gaps in
sampling.

\section{Practical Considerations}
\label{sec:practical}
In constructing the examples in \ref{sec:FvsLK}, some operational
details were kept hidden to minimize distraction. However
implementation of any period search algorithm  involves consideration
of frequency resolution, or making sure that the periodogram explores
all of the structures in the power spectrum, so that peaks in periodograms 
are not inadvertently skipped.

For non-regular or uneven time sampling, the usual Nyqvist criterion
does not apply. We posit that the criteria for ideal resolution follow
easily from a consideration of how the phases of observed points
change while stepping in frequency $f$, as discussed by
\citet{lafler67}.  If the separation in time between the first and
last observations is $T$, then at a putative period $P$ ($f = 1/P$)
the phase difference $\phi$ between the two points is given by
\begin{equation} \label{eqn:tau}
\phi = T . f  ~~-~~ M
\end{equation}
where M is an integer representing the number of full cycles elapsed
over the duration $T$. We want to take steps in $f$ that satisfy
\begin{equation}\label{eqn:dphi}
\Delta f = (\Delta \phi)_{max} / T
\end{equation}
where $(\Delta \phi)_{max}$ is the largest relative displacement in
phase between any two points produced by a step in the sampled
frequency. It is recommended that $(\Delta \phi)_{max} \leq 1/n $ 
where $n$ is the number of time-samples available, and $1/n$ is then
the mean spacing of observations in phase.

Let $T$ be the total duration of the observations. We want to sample
the power spectrum along a harmonic progression (equal intervals in $f
= 1/P$), for which a true Fourier series would have orthogonal
coefficients. Frequencies lower than $1/T$ are clearly undeterminable,
so we begin with the low value of $f$ as $1/T$. Using the result from
equation~\ref{eqn:dphi}, the $k^{th}$ frequency $f_{k}$ is thus given
by :
\begin{equation}
 f_{k} = 1/T + (k-1).(\Delta \phi)_{max} / T
\end{equation}
This progression of frequencies is then terminated at the highest
desired frequency (shortest period) to be tested.  The Lomb-Scargle
and Lafler-Kinman algorithms are both given this frequency array
over which the respective periodograms are to be evaluated.

\section{Application to Real Multi-band Data}
\label{sec:real_data}

We next derive $\Pi, ~\Theta$ and $\Psi$ for real measurements of a
real variable.  The data are from a multi-band study of variable stars
towards the Galactic bulge, using imaging data with the DECam
instrument \citep{flaugher15} on the Blanco 4m telescope at CTIO.
Multi-band (DECam bands $u, g, r, i$ and $z$) multi-epoch observations
of one of the variables are listed in Table~\ref{tab_B1_392}.  There are 58 
well measured  epochs in $u$, 68 in $g$, 
69 in $r$, 94 in $i$, and 81 in $z$. This
object, labeled B1-392 in the DECam study, was observed and reported
by \citet{soszynski14} from the OGLE project as OGLE-BLG-RRLYR-11078, and shown
to be an ab-type RR~Lyrae star with period $P=0.5016240$ days.  The OGLE
time coverage is much more extensive than for our DECam data, and
their derived period is expected to be very reliable.  The DECam data
include observations of a few nights in each of May, June and August
of 2013, followed by parts of 3 nights in February of 2015.  During
the 2013 nights, up to 4 observations were made on a night, covering a
span of up to 7 hours.  This is thus an excellent real world example
of sparse sampling infected with periodic patterns: diurnal, lunation,
and a nearly 2 year gap.  

\begin{deluxetable}{cccc} 
\tabletypesize{\scriptsize}
\tablewidth{0pt}
\tablecolumns{4}
\tablecaption{Multi-epoch Multi-band photometry of B1-392 \label{tab_B1_392} }
\tablehead{\colhead{HJD-2400000.0} &
  \colhead{Mag} &
  \colhead{$\sigma(mag)$} &
  \colhead{Passband} \\
  \colhead{} &
  \colhead{(instrumental)} &
  \colhead{}(rms) &
  \colhead{}   \\
}
\startdata
  56423.6596250  &   22.367  &   0.015  &   r  \\
  56423.6611280  &   21.957  &   0.014  &   i  \\
  56423.6626230  &   21.914  &   0.016  &   z  \\
  56423.6641170  &   23.789  &   0.015  &   g  \\
  56423.6656180  &   27.769  &   0.030  &   u  \\
  56423.8082430  &   22.623  &   0.015  &   r  \\
  56423.8097370  &   22.164  &   0.014  &   i  \\
  56423.8112550  &   22.129  &   0.019  &   z  \\
  56423.8127430  &   24.137  &   0.017  &   g  \\
  56423.8142220  &   28.160  &   0.029  &   u  \\
\enddata
\tablenotetext{a}{All magnitudes are instrumental, but share a common zero-point within each passband}
\tablecomments{Table~\ref{tab_B1_392} is published in its entirety in the electronic edition of the journal. A portion is shown here for guidance regarding its form and content.}
\end{deluxetable}
\begin{figure}[htb!]
\centering
\epsscale{0.85}
\plotone{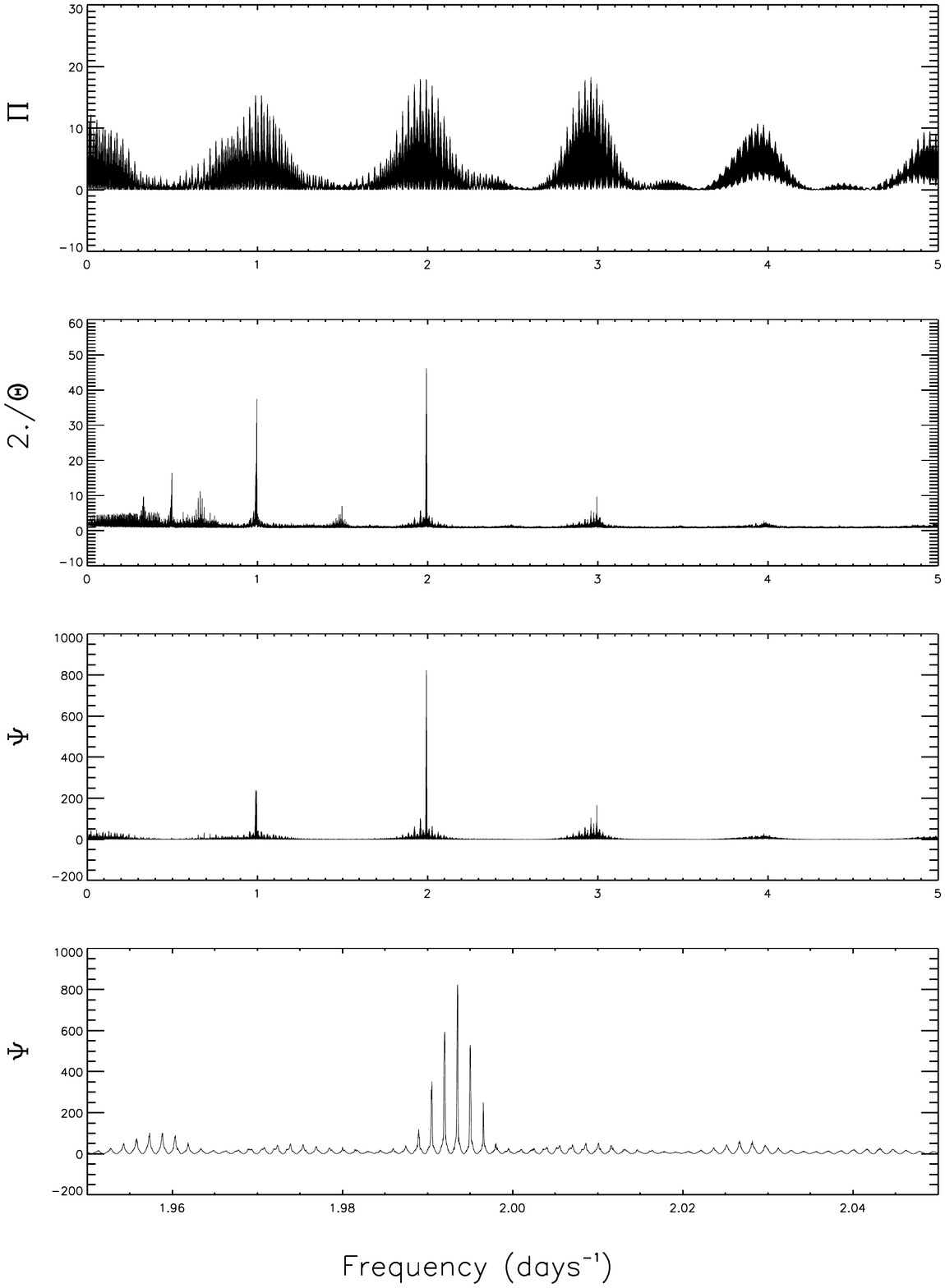}
\caption{The LS and LK periodgrams for the $g$ band observations in
  Table~\ref{tab_B1_392} are shown in the first and second panels
  respectively. The combined periodogram $\Psi$ is shown in the third
  panel, with an enlargement of the structure near $P \sim 0.5$ days
  in the bottom panel.  See \S~\ref{sec:real_data} for detailed
  explanation.}
\label{fig:392pgram}
\end{figure}

Figure~\ref{fig:392pgram} shows various
periodograms obtained from the $g$ band data alone.
The top panel shows the LS power spectrum, using $\Delta \phi_{max} =
0.02$ and searching for periods down to $P_{min} = 0.2$ days.  It is
impossible to decipher the period from this periodogram alone: it is
highly confused as a result of very sparse data with impressed
periodicities in the sampling and the multiple harmonics from the
saw-toothed light curve of the RR~Lyrae, all of which interact
mutually to redistribute the power into a noise-like forest of false
peaks. A peak near the period of $\sim 0.5$ days (or $f \sim 2$)
exists, but is not clearly distinguishable from the many aliases. The
second panel shows the LK periodogram, which is much better, but still
has multiple peaks.  The third panel shows the hybrid
periodogram $\Psi$ introduced in \S~\ref{sec:FvsLK}.  A prominent
peak shows at $f \approx 2$ ($ P \approx 0.5$) with relatively
weaker features also at $f \approx 1$ and $f \approx 3$, corresponding
to $P \approx 1$ and $P \approx 0.33$ days respectively. Examination
of the light curves resulting from these few putative frequencies
quickly shows that the real period is close to $0.5$ days.  The peaks
at $f \approx 1$ and $f \approx 3$ correspond to beat frequencies
$f_{b}$ of the true frequency $f_{0} \approx 2$, in accordance with
equation~\ref{eqn:beat}, due to the diurnal observing pattern.  These
are real aliases, and will express themselves through all algorithms.
The fourth panel is an enlargement of the third panel, showing how the
peak at $f \approx 2$ shows fine structure: the spacing between these
peaks corresponds to aliasing introduced by the $ \sim 600 $ days gap
in data between Aug. 2013 and Feb. 2015.  Symmetry arguments (since
each beat frequencies occur on both sides of the true period) and
examination of the light curve folded with each of the putative
periods indicated by the high peaks provides a good indication of the
true period. The highest as well as most central peak of the cluster 
corresponds to a period $P = 0.501628$ days, which is very close to 
the period reported by OGLE, and
within the precision that the two year observing window permits. The
lesson though, is that the $\Psi$ periodogram provides much cleaner
information than the $\Theta$ or $\Pi$ periodograms taken
individually.

\begin{figure}[htb!]
\centering
\epsscale{0.85}
\plotone{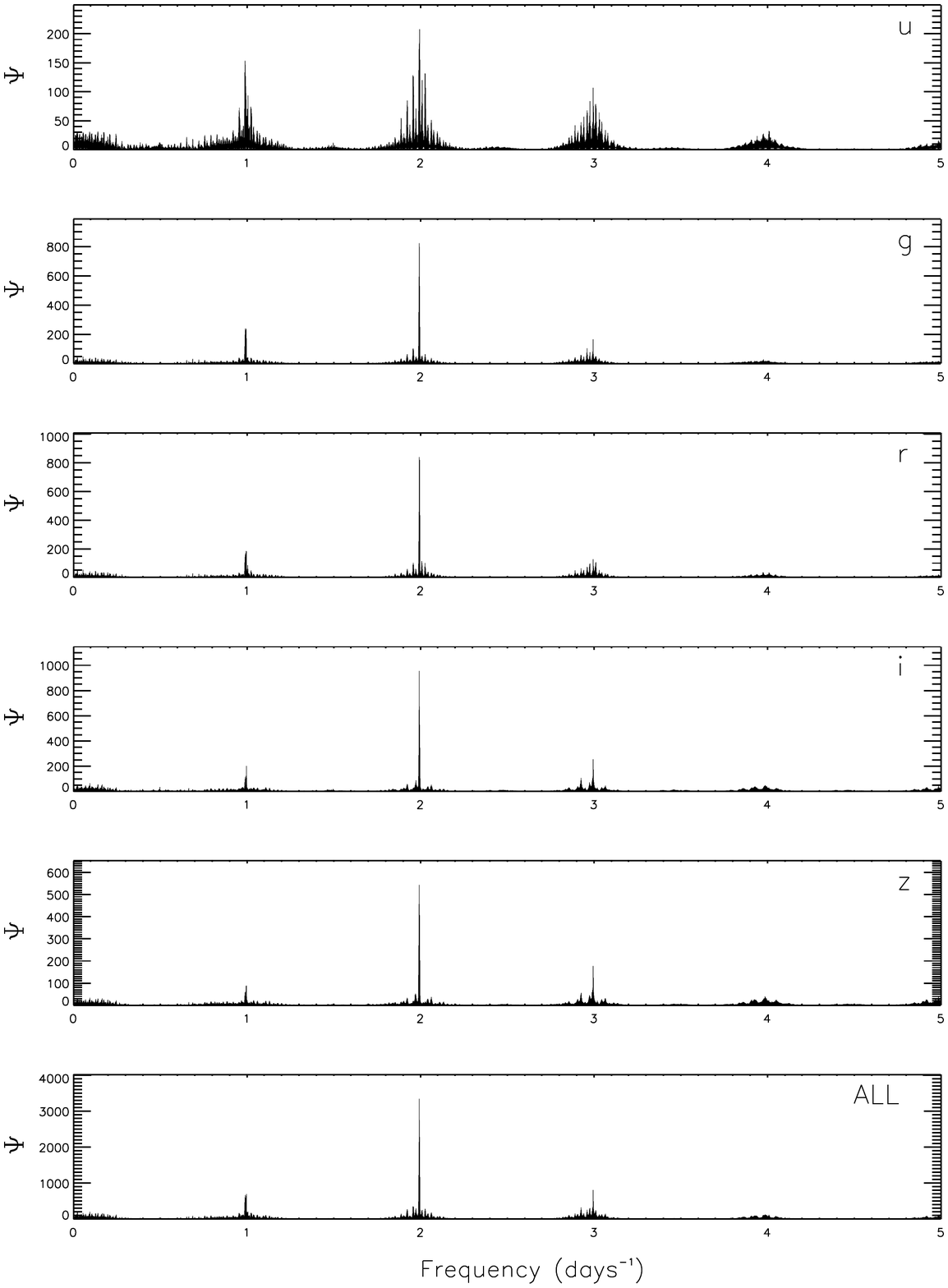}
\caption{The $\Psi$ periodogram for observations in each of the 5
  bands in Table~\ref{tab_B1_392} are shown in the top five panels.
  The bottom periodogram labeled `ALL' is the sum of $\Psi$'s for all
  bands.  By construction, the $\Psi$ statistic is agnostic about any
  phase information, and with the power in the harmonics due to light
  curve shape suppressed is very weakly dependent of how the light
  curves change from band to band. The average $\Psi$ is thus a useful
  way to aggregate information gathered at various epochs in different
  passbands, and so to derive the fundamental period using all of the
  data, as elaborated in \S~\ref{sec:real_data}. }
\label{fig:392psi_all}
\end{figure}

\begin{figure}[htb!]
\centering
\epsscale{0.85}
\plotone{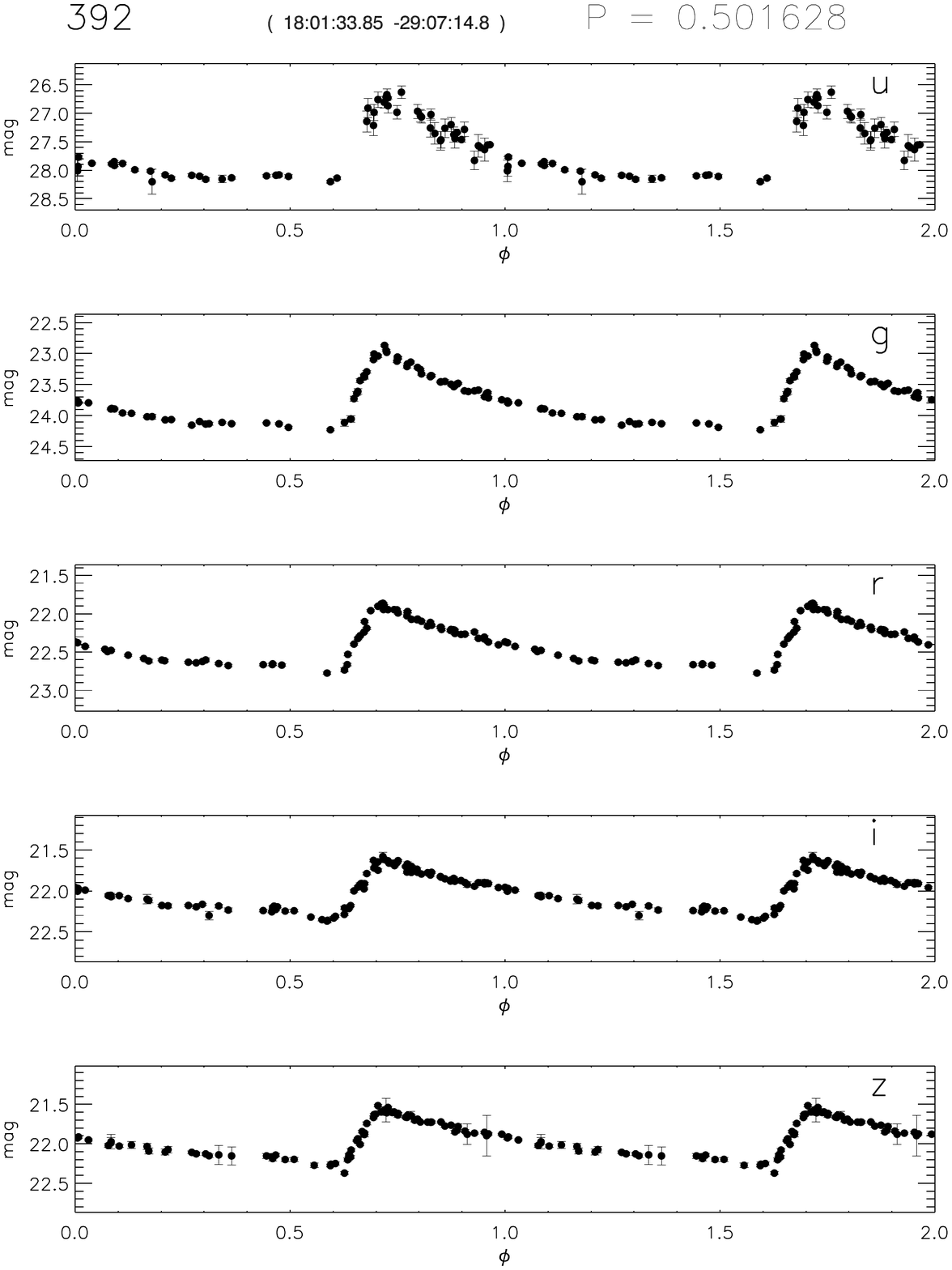}
\caption{The multi-band lights curves of B1-392 at $P=0.501628$, which
  is the frequency of the highest peak in the aggregate $\Psi$
  periodogram shown at the bottom of Fig.~\ref{fig:392psi_all}. The
  data are presented in Table~\ref{tab_B1_392}.}
\label{fig:392lc}
\end{figure}

Fig.~\ref{fig:392psi_all} shows the $\Psi$ periodogram in 5 passbands
($u,g,r,i,z$) derived from the data on B1-392 presented in
Table~\ref{tab_B1_392}.  The bottom periodogram labeled `ALL', is the
sum of all the periodograms from the individual passbands.
In this example the object was measured sequentially in all passbands in each visit, so that 
they are not really asynchronous measurements. 
Consequently all the same aliases show up in all 5
bands. However, had the sample times been different in each of the
bands, the aliases may have differed, though the peak from the true
fundamental period would be common.  The summed or co-averaged
$\Psi$ periodogram could then be an effective way of down-weighting specious
frequencies.  It could be argued that a different way of combining (e.g. taking the dot-product) 
the periodograms might be even more effective,  However in instances where one of the bands has very
little information, a more aggressive way of combining could be
counterproductive. For the time-being at least, we opt for simple
aggregation/co-averaging, which is the most robust against
pathologies.  $\Psi_{ALL}$ peaks at a frequency corresponding to $P = 0.501628$
days. Fig.~\ref{fig:392lc} shows the light curves in each of the 5
bands when folded with this period and co-phased to a common epoch
across the bands.

The $\Psi$ periodogram in the individual bands  is the product of the Lomb-Scargle power spectrum and the 
Lafler-Kinman statistic, neither of which retains any phase information.  Thus the summed $\Psi$ over the various bands is also expected to be independent of any phase and shape differences from band to band.  To verify that this assertion is really true, we ran the B1-392 example above by injecting phase differences in the light curves in the different bands.
Specifically, the observation times of all $g$ band epochs were retarded by 0.1 days, all $r$ band epochs by 0.2 days, all $i$ band epochs by 0.3 days, and all $z$ band epochs by 0.4 days.  The procedure was run on this altered data set, (taking care to keep the frequency sample points identical for all passbands). The resulting summed $\Psi_{ALL}$ spectrum was compared to that from the unaltered data: the difference was found to be {\it identically} zero at all sampled frequencies, demonstrating conclusively that $\Psi$ is indeed completely agnostic of phase differences.

\section{Estimation of Significance for a Reported Putative Period}
\label{sec:confthresh}

Consider the true signal $U(t)$, as in eqn~\ref{eqn:samp_window}. If
$U$ is constant in time, and is fully sampled, the LS periodogram (or
a Fourier transform) will yield a peak at zero frequency ($f = 0$),
and will be zero elsewhere. However, because of the missed samples,
and any temporal patterns in the sampling, power in the Fourier domain
will be distributed into various specious frequencies.  Conversely, we
can ask: how much specious power $\rho(f)$ can be put at frequency $f$
by a constant signal, due to gaps and patterns in the sampling?  The
relative power spectrum of $W(t_{i})$ ( i.e. $S(t_{i})$ with $U(t_{i})$
held constant at unity in eqn~\ref{eqn:samp_window}) provides the
answer.  To compare whether we have significant power $\Pi(f)$ from a
real measurement of $S(t_{i})$, it is useful to set $U$ to $\alpha$, the
amplitude of $S$, so that $\Pi$ and $\rho$ are
commensurate.  For us to be confident that periodicity reported at
frequency $f$ is significant, we would require that $\Pi(f) >>
\rho(f)$.  $\rho(f)$ also has another significance.  A sine or cosine
true signal $U$ with frequency $f_{0}$ which is perfectly sampled will
have a $\delta$-function for $\Pi$ at $f_{0}$, which will be
redistributed due to the incomplete and patterned sampling represented
by $W$ as $\rho(f - f_{0})$.  For a complex shape with many Fourier
components, the power at each component $f$ will be convolved by
$\rho$, resulting in a complicated $\Pi$ periodogram.  Conceptually one
should be able to use $\rho$ to deconvolve $\Pi$ (using their complex
rather than absolute value forms), but in practice this is an
ill-posed problem, especially in the presence of noise.
 
\begin{figure}[htb!]
\centering
\epsscale{0.85}
\plotone{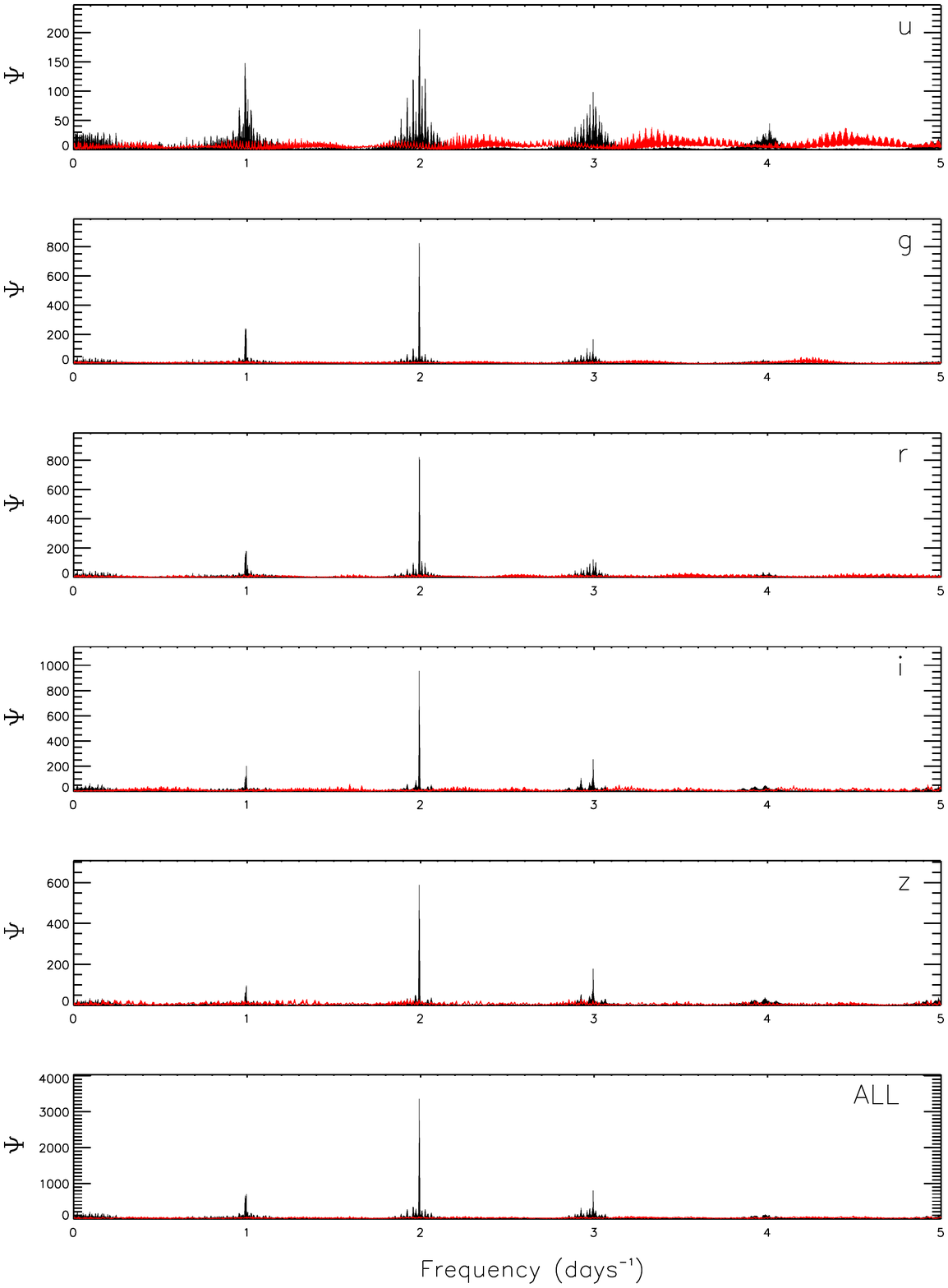}
\caption{Exactly the same at Fig.~\ref{fig:392psi_all}, but with the
  frequency dependent threshold $\psi_{thresh}$ for a significant ($\sim 1\sigma$)
  periodicity detection overplotted in red.  The confidence
  threshold calculation is discussed in \S~\ref{sec:confthresh}, which
  addresses both aliasing inherent in the actual sampling epochs, as
  well as the impact from noise from the photometry uncertainty
  estimates.}
\label{fig:392psi_conf}
\end{figure}

The effect of the presence of noise, and the significance of peaks in an LS periodogram 
is discussed in elegant detail by \citet{Horn86}.  Here we discuss the problem in a way that 
makes it easy to hybridize with the Lafler-Kinman method.
Let $\sigma(t_{j})$ be the estimated rms
uncertainty in the sampled value of $S(t_{j})$.  We make a random
drawing from a normal distribution with this $\sigma$, which we denote
by $\epsilon(t_{j})$.  We construct the quantity
\begin{equation}
S^{\prime}(t_{j}) = \alpha.W(t_{j}) + \epsilon(t_{j}) \\
\end{equation}
whose power spectrum $\rho^{\prime}(f)$ informs us whether a putative
periodicity detection at $\Pi(f)$ is significant or likely to be
specious.  While $\rho$ has been constructed above so that the total
power summed over all frequencies for $\rho$ and $\Pi$ are
the same, to assure commensurate comparison of $\rho^{\prime}$ and
$\Pi$, it is best to rescale $\rho^{\prime}$ by a constant so that
\begin{equation}
\sum_{f} \rho^{\prime}(f)  ~=~  \sum_{f} \Pi(f)
\end{equation} 

While $\rho$ and $\rho^{\prime}$ are well described mathematical forms
for evaluating significance of a putative frequency/period from FFT
analysis, the Lafler-Kinman periodogram $\Theta$ does not appear to
obey an easily characterizable mathematical behavior.  But if we are
to understand the significance of an indicated frequency/period from
the hybrid function $\Psi$, we must account for the contribution to
$\Theta$ from the aliasing. In what follows, we attempt to simulate 
functions analogous to $\rho$ and $\rho^{\prime}$ above that heed the
behavior restrictions for $\Theta$.

Unlike the FFT case, $\Theta$ for a noiseless steady signal is
indeterminate ($0/0$).  However, if we add the individual noise terms
$\epsilon(t_{j})$, we get deterministic values for $\Theta$ which
hover around $2.0$. Any particular such periodogram depends on the
particular random selection of $\epsilon$'s. Let us designate this
periodogram by $\eta(f)$.  The equivalent power $\mu(f)$ in the $\Psi$
domain from an unvarying entity in the presence of known estimates of
noise (as described above), is then given by:

\begin{equation}\label{eqn:conf1}
\mu(f)  = \rho^{\prime}(f) . \eta{f}
\end{equation}

$\Psi(f)$ should be greater than $\mu(f)$ for a putative period
indicated at $f$ to be significant.

We must also consider the case of a source which varies randomly and
aperiodically with an amplitude similar to that of the source under
consideration, i.e. $\alpha$ as used above.  For this we can take the
observations $[S,t]$ and scramble them by randomly matching the
observed quantity to the epoch.  Let the Lomb-Scargle and Lafler-Kinman
periodograms for this scrambled data-set be denoted by $\varpi(f)$ and
$\zeta(f)$ respectively.  By construction $\varpi(f)$ should be
co-normalized with $\Pi$, and $\zeta(f)$ is self normalizing.  For
this consideration, the discriminant for $\Psi$ is then given by:

\begin{equation}\label{eqn:conf2}
\xi(f) = \varpi(f) . \zeta(f)
\end{equation}

A conservative confidence threshold $\psi_{thresh}$ for the $\Psi$
periodogram can be obtained by co-adding the results for the two cases
above, so that

\begin{equation}\label{eqn:psi_thresh}
\psi_{thresh} (f)  ~=~  \mu(f) ~+~ \xi(f)
\end{equation}

Figure~\ref{fig:392psi_conf} shows $\psi_{thresh}$ plotted over the
$\Psi$ periodograms for our example object B1-392.

Note that neither $\mu(f)$ nor $\xi(f)$ shows the aliased `power' due to interaction of true periodicity with the observing window.  
As discussed above, this can be examined analytically for the 
FFT analysis by convolving $\rho$ with the power spectrum of the source, and depends on the full power spectrum (i.e the shape of the light curve)  of the source.  Since for our problem, the frequency 
and light curve shapes are unknowns, this is not directly useful. However, it may be possible 
put this property of $\rho$ to good use in an iterative fashion: once a putative period and shape 
are derived, it can be used to predict aliases,  which are then compare to the power spectrum derived from the observations. We do not develop this 
any farther in this paper, and must heed the caveat that $\psi_{thresh}$ \emph{does not} account for aliases 
resulting from interaction of the observing windows with the periodicity in the signal: such aliases will not be identified through $\psi_{thresh}$.

\section{Application to an LSST Simulation}
\label{sec:LSSTsim}

\begin{figure}[htb!]
\centering
\epsscale{0.85}
\plotone{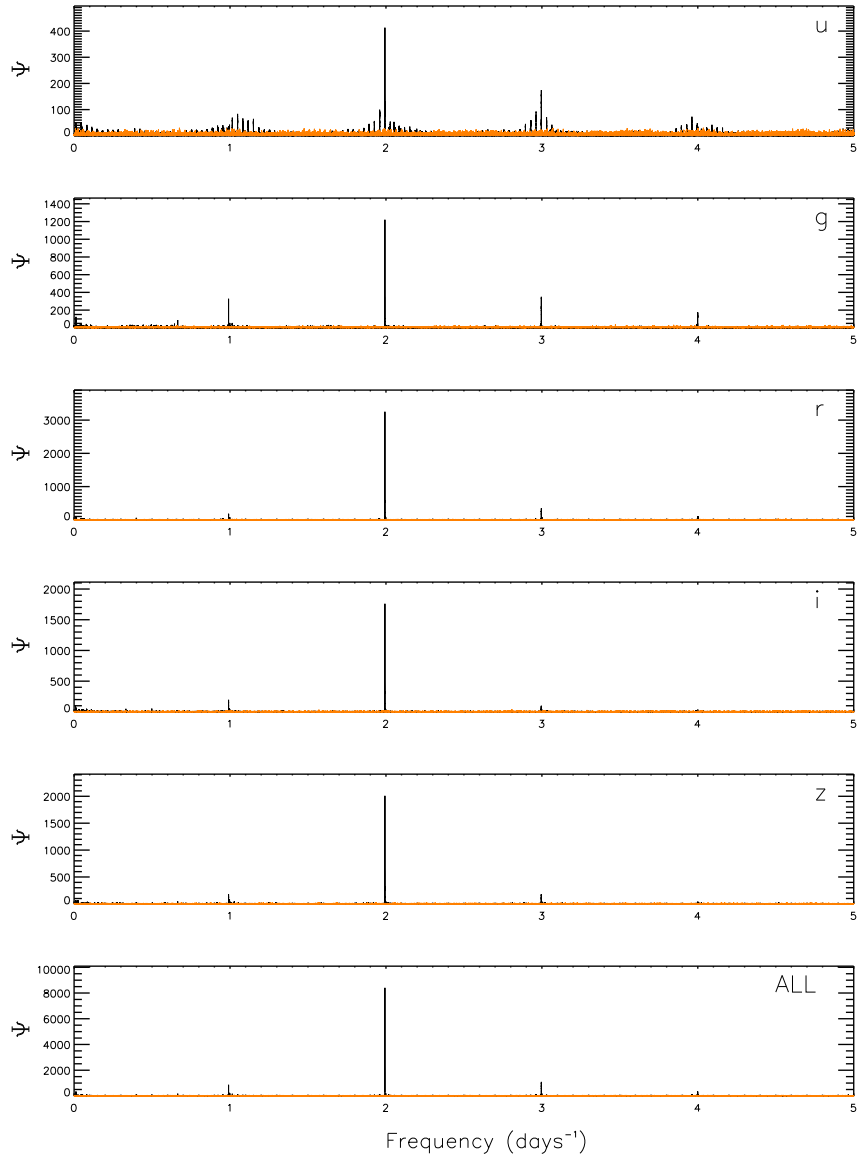}
\caption{The $\Psi$ periodograms (in black) and significance thresholds $\psi_{thresh}$ (in orange) for an entire LSST  10 year cadence simulation 
in $u,g,r,i,z$ for the light curve model for B1-392. The bottom panel shows the aggregate $\Psi$ and $\psi_{thresh}$ from all 5 bands.   }
\label{fig:10yrsim}
\end{figure}

In this final example, we consider a simulated cadence from LSST that samples the best fit template to the light curve and ephemerides
of B1-392.  We use the LSST Minion-1016 simulation, which is the baseline simulation (at the time of writing of this paper), and pretend that B1-392 
is always imaged as part of field 1256 ($\alpha \sim 293^{\degr}$ and $\delta \sim -32^{\degr}$) which is well observed in the simulation 
and outside the zone of Galactic avoidance in the projected survey. This example tests the proposition that the $\Psi$ statistics from the individual 
pass bands can be profitably combined even when the measurements in different bands are asynchronous.   There are 900 epochs for field 1256 spread over 10 years apportioned over 6 bands, of which we use only $u,g,r,i,z$, for direct comparison with the example with real data in \S~\ref{sec:real_data}. Noise drawn from a normal distribution with $\sigma = 0.03$ mag has been added to the simulated observations of the light curve model.  Fig~\ref{fig:10yrsim} shows the individual $\Psi$ spectra in each of the 5 bands, as well as the aggregate $\Psi$ taken collectively.  The true period at $P=0.501628$ days is clearly the dominant peak that appears near frequency $ f \sim 2 {~\rm days}^{-1} $. Satellite peaks at $f \sim 1, f \sim 3, f \sim 4$ are unavoidable, given the diurnal sampling window. In the $u$ band, where observations are restricted to within a few days of the new moon, satellite patterns due to a strong monthly window are also apparent.  Clearly, with all of the data from 10 years of observations in hand, derivation of the period is straightforward,  the diurnal and lunar sampling windows notwithstanding.

\begin{figure}[htb!]
\centering
\epsscale{0.85}
\plotone{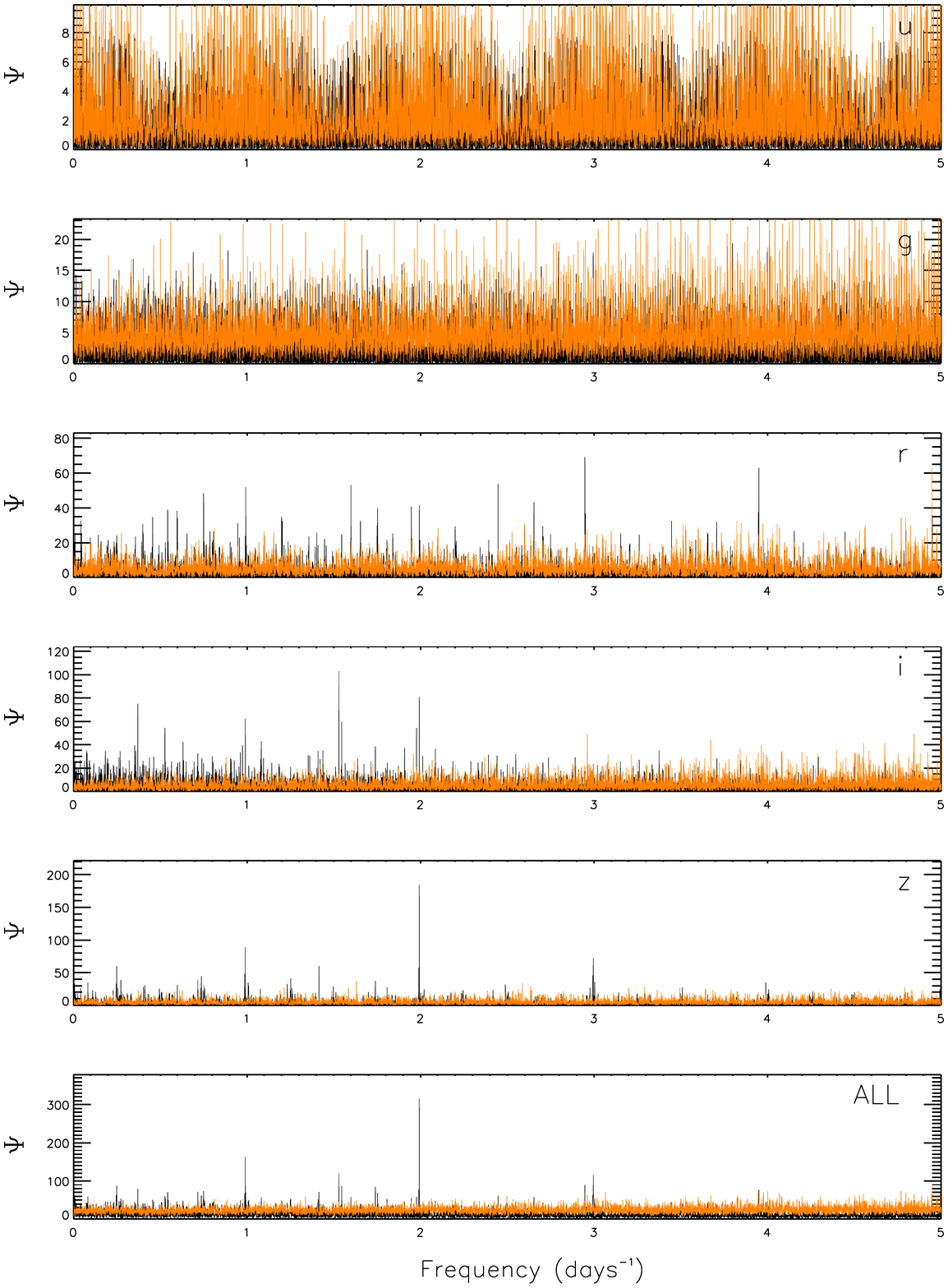}
\caption{Same as Fig~\ref{fig:10yrsim}, but using only the observations in the first 2 years of the simulated survey.}
\label{fig:2yrsim}
\end{figure}

Fig~\ref{fig:2yrsim} shows the results from only the first 2 years worth of observations. Note that in $u$ and $g$, there are too few observations, and $\psi_{thresh}$ dominates over $\Psi$.  In $r$, and even in $i$ there are strong false peaks. In $z$, which has the most observations, the true picture begins to emerge.  The bottom panel, which shows the aggregate of all bands, suppresses most of the false peaks: the true signal at $f \sim 2$ stands out, and weaker 
satellite peaks show up at the expected aliases.  One may therefore expect that objects like B1-392 could be characterized for periodicity after 2 years of LSST observations with the regular cadence in the Minion-1016 simulation.

\begin{figure}[htb!]
\centering
\epsscale{0.85}
\plotone{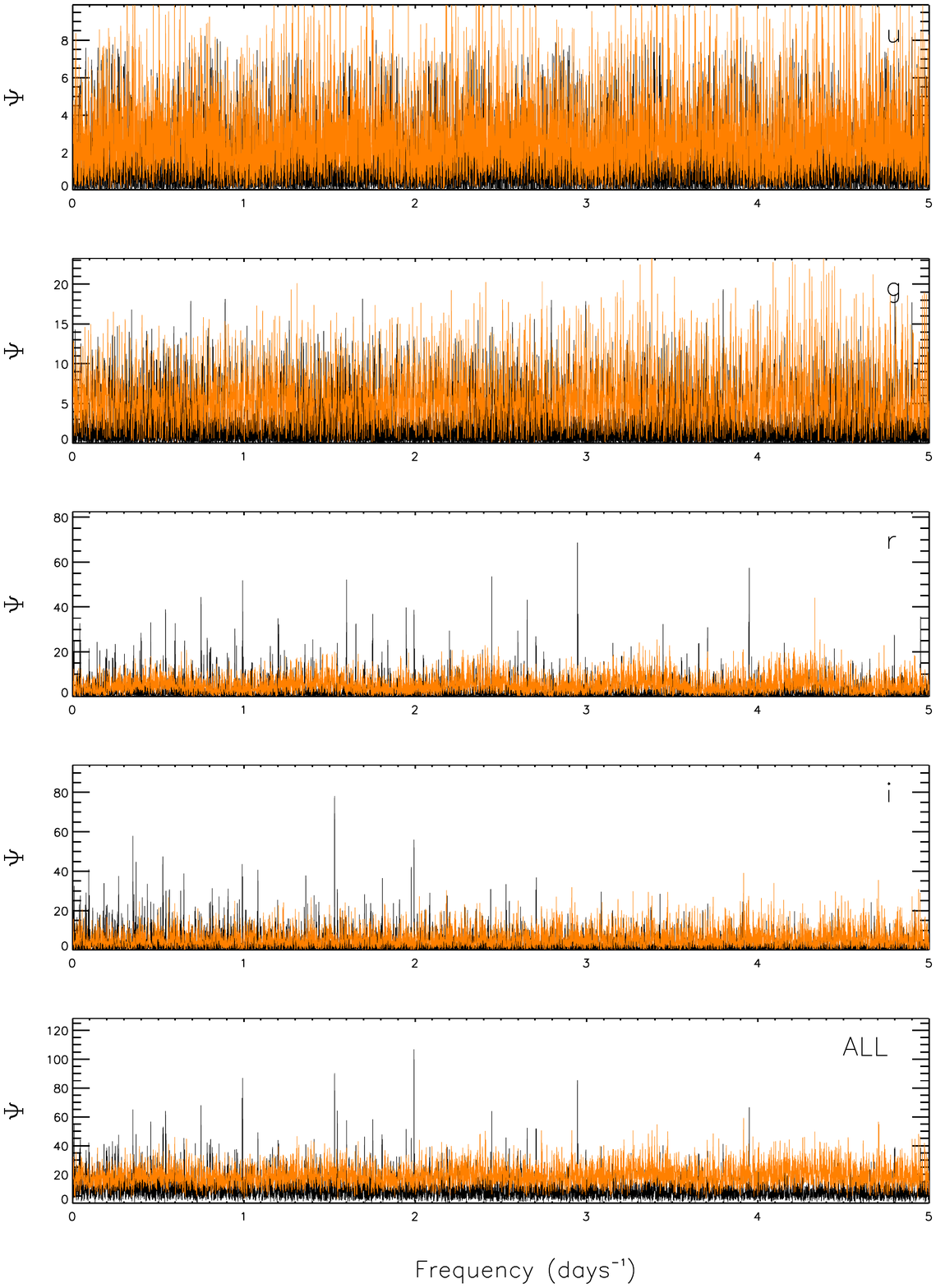}
\caption{Same as Fig~\ref{fig:2yrsim}, but using only the $ugri$ data. Note that while the correct period is not identifiable in any of the individual bands, in the summed $\Psi$ periodogram it begins to emerge as the tallest peak (albeit barely).} 
\label{fig:2yrsimnoz}
\end{figure}

Since the final $\Psi$ spectrum in this example is dominated by the $z$-band contribution, it is interesting to examine 
Fig~\ref{fig:2yrsimnoz}, where the above case has been re-run without the $z$-band data.  None of the individual bands indicate the correct period by themselves, but the tallest peak in the co-added $\Psi$ periodogram corresponds to the 
correct value.

\section{Concluding Remarks}
\label{sec:Conc}

The method presented here has been developed and used to
examine/analyze light curves of over 20,000 putative variables from
time-domain images of select fields in the Galactic bulge obtained
with DECam on the 4-m Blanco telescope at CTIO. The example 
in \S \ref{sec:real_data} for the
object B1-392 was one such object.  This procedure was used
to generate putative light curves of all of these objects, simply
by identifying the maximum value in their respective multi-band
co-added $\Psi$ periodograms.  Objects of interest can be selected
from these `initial' light-curves, and periods and light curves
refined through interactive examination of any aliases present in the
periodograms.  Our experience has been that at least for the data at
hand and the range of periods of primary interest (RR Lyrae stars),
the maximum value co-added $\Psi$ is the best estimable period in 95\%
of the cases. We have also compared our derived periods for $\sim 2000$ 
RR Lyrae stars with those in common with OGLE.  OGLE derived
periods from a single passband, but had a much larger number of sample
epochs, and better cadence.  As for the case of B1-392, we find over
98\% of the periods independently derived by the method presented here
are indistinguishable from those reported by OGLE within the precision
permitted by our 2-year observational baseline.

An earlier implementation of $\Psi$ utilized a Fast Fourier Transform
(FFT) instead of the Lomb-Scargle algorithm. While the former more
closely adheres to the underlying assumptions of a Fourier series, it
requires laying the observational timeline onto a regularly spaced
temporal manifold, where time sample points with missing data are
assigned zero-value, in accordance with eqn.~\ref{eqn:samp_window}.
Because the FFT produces a frequency spectrum where the lowest
frequency is the inverse of the total time duration, to get the
frequency series that adequately samples phase intervals $(\Delta
\phi)_{max}$ (as utilized in eqn.~\ref{eqn:dphi}), the re-sampled time
sequence must be extended to a total duration $\tau$, where $\tau = T
/ (\Delta \phi)_{max}$.  This is typically several tens of times
longer than the actual observing window.  Moreover, if we want to
probe all periods $P > P_{min}$, then the re-sampled sequence must
contain time steps $ < < P_{min}$. For short values of $P_{min}$ and
large $\tau$, the re-sampled array becomes extremely large, and mostly
full of padded zero values. While interpolation can be avoided by
treating each sample point as a bin, one must take care of any
situations where there are multiple observations within any time bin
(by error weighted co-averaging).  In practice, the construction of
this artificially large re-sampled time series is inefficient.  The
$\Psi$ periodograms that result from the FFT vs the LS approaches have
no significant differences.  While it may be of academic interest how
closely the LS algorithm mimics a real Fourier series, (see \citet{vanderP17} 
for a pertinent discussion) at the end of
the day the worth of a period finding procedure is in its empirical
efficiency.  The examples discussed also show how efficient the
Lafler-Kinman algorithm really is for the cases where the light curve
departs significantly from sinusoidal (albeit by visualizing it as the
inverse of $\Theta$).  

In the example in \S~\ref{sec:real_data} for 
the observed data of B1-392, the periodograms were computed for 
$\sim 173000$ frequency points ($(\Delta\phi)_{max} = 0.02$). The 
compute time to generate the $\Psi$ periodograms in the $ugriz$ bands 
on a MacBook Pro were 2.95, 3.29, 3.27, 4.07 and 3.64 
seconds respectively, totaling to 17.3 seconds for the co-added $\Psi_{ALL}$.
The time taken scales linearly with the number of frequency points.  Not much 
would be lost had we run with ($(\Delta\phi)_{max} = 0.05$), which would 
cut the run times by a factor of 2.5.  The implementation of this case posted on github
(see Appendix) also calculates the significance thresholds $\psi_{thresh}$, which involves 
calculation of the two additional $\Psi$ like spectra $\mu$ and $\xi$ (see \S~\ref{sec:confthresh}). This 
triples the compute time to run that example.
 
The multi-band periodogram method of \citet{vanderP15} has addressed
the issue of combining asynchronously sampled multi-band time sequence
measurements.  Theirs is an extension of the Lomb-Scargle algorithm,
where they seek to construct a common mode light curve that contains
only low-order harmonics, and where the individual channels are
characterized by departures from the common mode variation. Key to
their procedure is a mechanism that picks out the low-order Fourier
terms that are common to all channels.  \citet{mondrik15}  profer a multi-band 
extension of the analysis of variance (AoV) periodogram, which is another 
variant of harmonic analysis that is generalized by them for multi-band use.
By contrast, in the procedure
presented here the PFM based analysis is used to marginalize the
higher harmonics from the harmonic analysis (and vice-versa for the
sub-harmonics that appear in the PFM analysis) which effectively
isolates the fundamental frequency alone that is common to all
channels.  

While not the objective of this paper, a performance
comparison of these two methods on simulated LSST cadences for various
object classes and period ranges is clearly of interest.  
In a very cursory effort we have also run our procedure on a posted 
example\footnote{We modified the Jupyter notebook example in 
\url{http://nbviewer.jupyter.org/github/astroML/gatspy/blob/master/examples/MultiBand.ipynb}  
to run locally in our machine and have local access to the data} of the \citet{vanderP15} algorithm, 
which uses the multi-band light curve of RR Lyrae \#1013184 in SDSS Stripe 82 \citep{sesar10}.  
The run times to produce the $\Psi$ periodograms for this example with ∼ 67000 
frequency sample points were about 1.2 seconds in each of the 5 passbands, 
totaling ∼ 6 seconds to obtain $\Psi_{ALL}$. Our run with the {\it gatspy} \citet{vanderP15}  
code on this example, with a similar number of sampled frequencies 
ran in a little under  3 seconds. While there are many details to 
sort out to make a like-to-like comparison, it appears that they both run on 
comparable time-scales.

\acknowledgments

This work has its roots in conversations with Gautham Narayan, who has
been engaged in employing period finding methods on existing survey
data in the context of building algorithms for the transient event
broker ANTARES \citep{saha16}.  We thank the referee for perceptive and useful comments.

\software{Psearch \citep{Saha17}}

\appendix

\section{Code for Processes and Algorithms in the IDL Language}

The algorithms discussed in this paper are coded in the IDL language, and are available through {\it github} at {\it https://github.com/AbhijitSaha/Psearch }
and Zenodo \citep{Saha17}. The data for the RR~Lyrae stars B1-392, as well as for the LSST year simulation used in this paper are also available from that location. Descriptions and instructions for use are given in comments within the routines, and/or in the accompanying {\it README} file. Also included is the {\it scargle.pro} routine from the NASA IDL astronomy library. Two examples of how to use the code are also provided, which works with a) the B1-392 data-set that is provided, and b) the simulation of B1-392 as it might be observed by LSST as it is discussed in this paper.

\bibliographystyle{aasjournal}
\bibliography{ms}

\end{document}